# A model of the vicious cycle of a bus line

Asaf Bar-Yosef[1], Karel Martens[2], Itzhak Benenson[1,$]

[1]Department of Geography and Human Environment, Tel Aviv University, Israel

[2]Institute for Management Research, Radboud University Nijmegen, the Netherlands

asafbj@yahoo.com, k.martens@fm.ru.nl, bennya@post.tau.ac.il

[$]Corresponding author

## Abstract

Several authors have noted that in a non-regulated environment the development of public transport service is self-adjusting: Faced with decreasing demand, operators will tend to reduce service to cut costs, resulting in a decrease in the level-of-service, which then triggers a further drop in demand. The opposite may also occur: high demand will induce the operator to increase supply, e.g. through an increase in frequency, which results in a higher level-of-service and a subsequent increase in passenger numbers, triggering another round of service improvements. This paper adds to the literature by presenting an analytic model for analyzing these phenomena of vicious or virtuous cycles. The model formalizes passengers' decisions to use a public transport service depending on waiting time and employs field data regarding passengers' variation in willingness-to-wait for a public transport service. The paper investigates the dynamics of the line service and shows how the emergence of a vicious or virtuous cycle depends on total number of potential passengers and share of captive riders. It ends with a discussion of the implications of the findings for the planning of public transport services.

# 1. What is vicious cycle?

When a bus line operator is faced with low demand, he tends to cut expenses instead of improving level-of-service (LOS) (Reinhold, 2008). The easiest way to do that is to reduce the frequency of buses. In response, some of the passengers may use the service less or not at all, as they change transportation mode, destination choice, or decide to forego a trip altogether (Castaline, 1980; Bly, 1987).[1] This decrease in the demand for the bus service may cause a further decrease in bus frequency. This vicious cycle continues until only captive passengers, who do not have an alternative mode of travel, continue using the line (Downs, 1962).

The positive, virtuous cycle, is also observed: high demand for a bus line induces the operator to supply more buses and this causes an increase in LOS and perceived satisfaction of passengers. Non-captive passengers react to the improvement, switch from private cars to buses (Liu et al., 2010), leading to a rise in demand (Kingham et al., 2001), which in turn triggers a further increase in LOS.

The above negative and positive cycles are well known among transportation researchers and have been investigated both theoretically (Martens and Hurvitz, 2011) and in practice, as in Xu et al. (2010), who mention in their review of the history of the public transportation network in Beijing that poor public transportation systems may lead to lines' vicious cycle. In contrast, Reinhold (2008), in a study of Berlin bus network, shows that a virtuous cycle can be induced, even when facing a declining revenue stream from passengers. By deliberately increasing the frequency on a limited number of main bus lines, while cutting service on supplementary lines to reduce total operating costs, in combination with improvements in communication and marketing, the Berlin bus operator succeeded to boost ridership on lines with frequency increases, resulting in a substantial rise in overall passenger volume and revenues.

Levinson and Krizek (2008) were the first who tried to investigate the vicious cycle analytically. They introduced three hypothetical dependencies: one linear, of bus speed as a function of bus waiting time, and two non-linear, of the number of passengers as a function of bus speed, and of waiting time as a function of the number of passengers that use the line. The non-linear dependencies resulted in a qualitative conclusion that the bus line has two stable states – one in which the bus waiting time is very high and the number of passengers is close to zero, and the other, in which the waiting time is close to zero and all potential riders use transit. At the same time, the dependencies they used are purely hypothetical and the authors do not propose any mechanisms that can explain them. As a result, the number of passengers in an equilibrium state is unrealistically high or low (see Appendix C for more details).

---

[1] Passengers may indeed respond in all these ways to a change in service frequency. For reasons of readability only, in the remainder of the paper we will limit these possible behavioral responses to mode change among non-captive travelers. This limitation has no consequences for the results presented in the paper.

In what follows, we construct a model of the vicious cycle phenomenon as an outcome of passengers' willingness-to-wait for a bus. The model is based on the assumption that the bus operator is interested to maximize its profits (Van Nes, 2002). Based on experimental data on willingness-to-wait for a bus, we develop a full analytical model of the interaction between the passengers and the bus company, and formulate the conditions when the negative (vicious) or positive (virtuous) cycles emerge. Following the tradition, we call the phenomenon "vicious cycle" and the model "the vicious cycle model". The aim of the paper is to provide a systematic understanding of the dynamics of the vicious cycle under various circumstances.

The paper is organized as follows. In Section 2, we present our assumption regarding the distribution of passengers' willingness-to-wait for a bus, based on a brief overview of surveys into the issue. Then, in section 3, we analyze the interplay between passengers' willingness-to-wait and public transport frequency. Based on this, we present the dynamics of the vicious cycle as dependent on total number of potential passengers and share of captive riders (Section 4). Section 5 applies the model to the problems of bus size and the introduction of a new line. We end with a discussion in which we elaborate on the relevance of the findings for real-life public transport service and suggest some directions for further refinement of the model (Section 6).

## 2. Field estimates of passengers' willingness-to-wait for a bus

Public transport ridership depends on a diversity of factors, such as in-vehicle travel time, walking distance to a bus stop, walkability of the urban environment, public transport fares, as well as the quality and cost of alternative modes of transport. Among these factors, waiting time is particularly important, as riders view waiting time as much more burdensome than an equivalent amount of time spent in travel (Ceder, 2007). Conventional wisdom holds that average waiting time equals one half of the expected headway (the time gap between two consecutive buses) (Hess et al., 2004). While passengers can reduce their waiting time by synchronizing their arrival at a public transport stop with the service schedule, this strategy is only effective if transit service is reliable. Moreover, the strategy often only replaces waiting time at the stop with hidden waiting time at the origin of a trip (i.e., at home) (Furth and Muller, 2006).

The willingness-to-wait for a bus or other public transport service varies among passengers, as has been shown in various surveys performed over the past two decades estimating the distribution of the maximal bus waiting time $\tau$ that passengers are ready to accept.

Peterson et al. (2006), in a stated preference survey, studied how long bus users are willing to wait for a free transfer between two bus operators. They obtained that only 10% of the respondents indicated to be willing to wait for more than 15 minutes and only 3% for more than 20 minutes. Kim and Ceder (2006) asked potential passengers if they would use a planned shuttle service as depending on the time interval between buses. Half of the passengers responded that they definitely or probably accept a 5 or a 10 minutes waiting time, while more than 70% raised doubts that they would wait for 15 minutes or more. In a city-wide survey conducted in Dublin, 90% of the respondents claimed that they are not willing to wait for more than 20 minutes, and 70% were not willing to wait for more than 10 minutes (Caulfield and O'Mahony, 2009). Only 2% claimed they would be willing to wait more than 30 minutes.

Since stated preference studies do not necessarily accurately reflect people's revealed preferences, it is worthwhile to compare these findings with a revealed preference survey. In such a survey, Hess et al. (2004) studied the transport behavior of college students who had the choice between two identical bus lines serving the same origin-destination pair, one line served by blue buses and the other by green buses. Students could ride for free on the 'blue' bus (with the university paying for the fare on behalf of the student), but had to pay a fare of 0.75 US $ for a ride on the 'green' bus. The headway between buses on the blue line was on average 10 minutes; between green buses, it was 12 minutes. Hess et al. (2004) measured the waiting time of the students who decided to wait for a blue bus rather than board a green bus when the latter arrived first at the bus stop. They found that the average additional waiting time for these riders was 5.8 minutes (with an STD of 3.3 minutes); the median elapsed time was 4.5 minutes. With a headway of 10 minutes between the blue buses, these findings are largely in line with the conventional wisdom that actual waiting time equals, on average, one-half of the expected headway. Since virtually all students used the bus line on a regular basis, they will have had a relatively accurate estimate of the expected additional waiting time when deciding not to board an available green bus. The findings can thus be interpreted as students' actual willingness-to-wait for a bus. Important for the purposes of this paper, is the finding that the distribution of willingness-to-wait time is comparable to those found in stated preference studies (see below). Hence, we conclude that stated preference findings can be used for the analysis of the vicious cycle phenomenon.

Figure 1 presents the results of the two stated preference surveys of passengers' willingness-to-wait that are employed in our model. The results have been derived from the studies of Peterson et al. (2006) and Caulfield and O'Mahony (2009). We approximate these empirical distribution densities by the function $f(\tau) = C\tau^a e^{-b\tau}$, with the values of parameters a and b estimated using the non-linear regression method of SPSS19 and C serving as a normalizing constant for ensuring $\int_0^{30} f(\tau) = 1$.

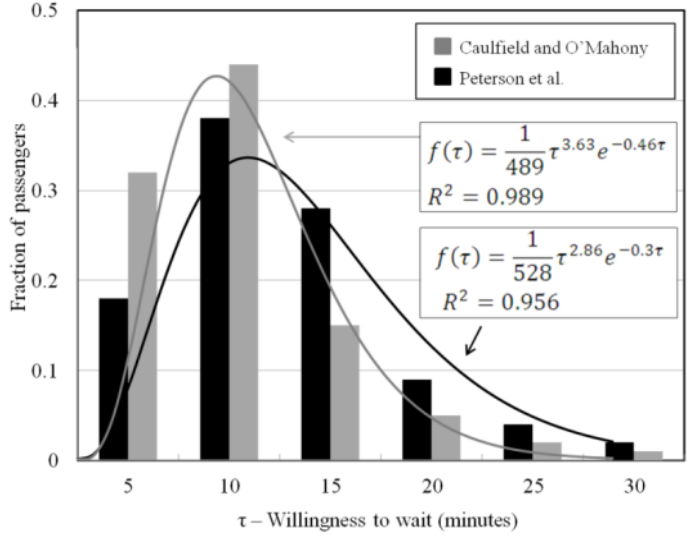

Figure 1: Experimental distributions of passengers' willingness-to-wait time $\tau$, approximated by the function $f(\tau) = C\tau^a e^{-b\tau}$.

The reaction of the passenger population to waiting time and, hence, to bus frequency, is crucial for understanding modal shift between car and bus (Kingham et al., 2001) or, more generally speaking, for understanding bus use. Hence, it is a key element in the model of the vicious cycle and in what follows we perform a deep analytical investigation of this phenomenon. We intentionally ignore other factors that influence the usage of public transport, such as accessibility of destinations, comfort of the service, car travel times, or parking availability and costs. These factors obviously will have an impact on the occurrence of vicious and virtuous cycles in a real-world setting, but do not change the fundamental dynamics of the phenomenon.

### 3. From willingness-to-wait to bus line dynamics

The model considers a circular bus line of a trip time L. Let the entire population that can be served by the line during this trip be indicated by $P_{total}$, the fraction of captive passengers by g, and the fraction of non-captive passengers by 1 – g, i.e. the line population consists of $gP_{total}$ captive passengers and of $N = (1 - g)P_{total}$ non-captive passengers. Captive passengers always make the trip, while non-captive passengers decide on using a bus depending on the time $\tau$ they have to wait at a stop. Let us denote as N($\tau$) the number of non-captive passengers whose maximal waiting time is between $\tau$ and $\tau + d\tau$, that is, $N = \int_0^\infty N(\tau)d\tau$.

Let us consider the dynamics of the passengers taking the line at a daily time resolution. At a given day d, some non-captive passengers who waited for the bus that day longer than their maximal waiting time $\tau$, may decide to stop using the bus line in the future. We assume that this decision is made after several failures and, thus, the fraction of those who left at a day d is lower than the fraction of those who waited for too long that day. In parallel, we assume that some of the non-captive passengers, who currently do not use the bus, attempt to use it in the hope that the waiting time is reduced in comparison to previous experiences. Some of them may succeed to board several times in a row after waiting less than τ and may thus change their mode back to the bus.

Let B(d) be the number of buses serving the line at a day d and $T_{B(d)}$ be the time interval between buses ($T_{B(d)} = L/B(d)$). Let us further assume that every passenger uses the bus once a day and that the daily rate at which passengers switch from being a user to becoming a non-user is β, while the rate of the opposite switch is α. Users for which $τ > T_{B(d)}$ are all served, while users with $τ < T_{B(d)}$ are served if they arrive to the stop τ minutes or less before the bus arrival. Assuming that the passengers arrive to the bus stop randomly in time, the fraction of the served users among those of $τ < T_{B(d)}$ is $τ/T_{B(d)}$.

Let us denote the number of non-captive passengers whose maximal waiting time is τ and who use the bus at a day d as $N_u(τ, d)$, and those who do not use bus at a day d as $N_n(τ, d)$, $N_u(τ, d) + N_n(τ, d) = N(τ)$.

The dynamics of $N_u(τ, d)$ and $N_n(τ, d)$ can be presented as follows:

If τ < $T_B$ then

$$N_u(τ, d+1) = N_u(τ, d) + αN_n(τ, d)\frac{τ}{T_B} - βN_u(τ, d)\left(1 - \frac{τ}{T_B}\right) \quad (1)$$

$$N_n(τ, d+1) = N_n(τ, d) - αN_n(τ, d)\frac{τ}{T_B} + βN_u(τ, d)\left(1 - \frac{τ}{T_B}\right)$$

Otherwise

$$N_u(τ, d+1) = N_u(τ, d) + αN_n(τ, d)$$
$$N_n(τ, d+1) = N_n(τ, d) - αN_n(τ, d)$$

The number P(d) of passengers served at a day d is given by:

$$P(d) = gP_{total} + (1-g)\int_0^∞ Nu(τ, d)dτ \quad (2)$$

To proceed, let us assume that a bus company decides on bus frequency once in a quarter of a year, and that β is essentially higher than α. That is, passengers are more inclined to stop using a service if the bus does not arrive during their maximal waiting time, than to resume using the service in a hope that waiting time has decreased. Figure 2 presents the distribution of the passengers who continue to use the bus line (dark grey) versus the initial distribution of the passengers by their maximal waiting time, at d = 0 (light and dark grey together), and after a quarter of a year (d = 90 days), for $T_B$ = 15 min and different values of α and β.

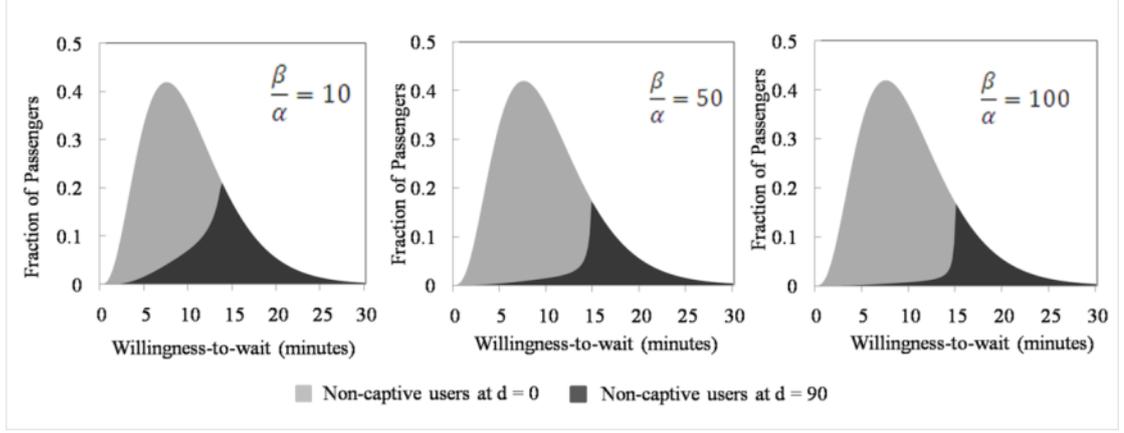

Figure 2: The distribution of $N_u(\tau, d)$ at day d = 0 (light grey) and at day d = 90 (dark grey), for α = 0.005, β = 0.05, 0.25 and 0.5 (all per day), $T_B$ = 15 minutes and the distribution of willingness-to-wait according to Peterson et al. (2006), assuming all non-captive passengers attempt to use the bus service from day d = 0 onwards.

According to Figure 2, during a three month period the majority of passengers with τ < $T_B$ = 15 leave the line in case β/α is sufficiently large.

Below we assume that the bus company can change the bus frequency once in a quarter and, for simplicity, ignore passengers with τ < $T_B$ that continue to use the bus after the end of the quarter. In this case, $\int_0^\infty N_u(\tau, d)d\tau$ can be presented as $N \int_{T_{B(d)}}^\infty f(\tau)d\tau$, and the dynamics of the bus users, by quarters $t$, can be presented as:

$$P(t+1) = gP_{total} + (1-g)P_{total} \int_{T_{B(t)}}^\infty f(\tau)d\tau, \qquad (3)$$

where, as defined above, $P_{total}$ is the overall number of potential passengers, g is the fraction of captive passengers among them, and t denotes a quarter of a year.

The full model, which accounts for passengers with τ < $T_{B(d)}$ who continue to use the bus line, is presented in Appendix B, where we demonstrate that the dynamics of (3) and of the full model are qualitatively similar.

For the next step of the analysis, let us substitute, in (3), $T_B$ by L/B(t), and denote by m the *optimal number of passengers* per bus that makes it maximally profitable for the operator to run the line. The value of m evidently depends on the maximal capacity of a bus, but will be somewhat lower than that to avoid overcrowded buses (and thus a risk that potential passengers will have to be left behind due to a lack of capacity), but higher than the number of passengers necessary to merely cover operation costs (Meignan et al. 2007). Equation (3) can be thus finalized as

$$P(t+1) = gP_{total} + (1-g)P_{total} \int_{\frac{Lm}{P(t)}}^\infty f(\tau)d\tau \qquad (4)$$

Assuming that the number of buses is a continuous variable and denoting $P_{total}/m$ as $R_{total}$, equation (4) can be transformed into the equation of bus dynamics by quarters of a year:

$$B(t+1) = F(B(t)) = gR_{total} + (1-g)R_{total} \int_{\frac{L}{B(t)}}^{\infty} f(\tau)d\tau \qquad (5)$$

Where $B(t) = P(t)/m$, and $gR_{total} = gP_{total}/m$ and $(1-g)R_{total} = (1-g)P_{total}/m$ express the numbers of captive and non-captive passengers, respectively, in terms of the number of buses necessary to carry them. Equation (5) thus represents the dynamics of a bus line's vicious cycle through the dynamics of the number of buses on the line.

## 4. Study of the vicious cycle

The dynamics of (5) are defined by the shape of *F(B(t))* (Holmgren, 1996); let us investigate it as dependent on $R_{total}$ and g.

*F(B(t)) monotonously grows with the growth of B(t):* F(B(t)) is an integral of a positive function and with the growth of B(t) the lower limit of the integral decreases.

*Asymptote:* With the growth of B(t), the lower limit of the integral in (5) tends to zero, the integral tends to 1, and, thus, F(B) → $R_{total}$.

*Equilibria:* Equilibrium number of buses B*, if it exists, satisfies the equation B* = F(B*). Equilibrium B* is locally stable if $|F'(B^*)| < 1$ and unstable if $|F'(B^*)| > 1$ (Holmgren, 1996). For monotonously growing $F(B)$, $F'(B)$ is always positive and, thus, the conditions of local stability can be simplified: B* is stable if $F'(B^*) < 1$, and unstable, if $F'(B^*) > 1$. Below, we omit the case of $F'(B^*) > 1$.

Analytical investigation of monotonously decreasing willingness-to-wait $f(\tau)$ is presented in Appendix A: equation (5) has one or three equilibria depending on $R_{total}$ and g and on whether $f(\tau)$ decreases, with τ, faster or slower than $1/\tau^2$ in this case.

For a real-world non-monotonous distribution of the willingness-to-wait $f(\tau)$, the possible number of equilibria of (5) is also one or three. To investigate the stability of these equilibria, let us note that they are the points of intersection of the curve B(t+1) = F(B(t)) and of the straight line B(t+1) = B(t). Stability of the equilibria is thus defined by the tangent to F(B(t)) at the point of intersection.

In Figure 3, four qualitatively different situations are represented regarding the equilibria of (5) as dependent on $R_{total}$ and g. They are all highly intuitive: in case the population ($R_{total}$), as defined in terms of number of buses, is low (Figures 3a, 3b), the vicious cycle is inevitable. No matter how high the initial number of buses, the overall number of passengers is insufficient to (financially) justify the line and the bus operator will decrease bus frequency until only captive passengers will remain on the bus. Note that in case of high operating costs of the bus line, no equilibrium may actually exist, as the number of captive riders may be too low to warrant any bus operations.

The opposite case is represented by a large population ($R_{total}$) and a high share of captive riders g (Figure 3d). This situation is obviously the best for the bus operator, as a virtuous cycle is inevitable in this case: the high number of captive passengers will justify an initially high bus frequency that will automatically attract non-captive passengers, which will induce the operator to further increase bus frequency. As a result, the majority of the population will ultimately use the bus line.

The most interesting is the case when the number of potential passengers is high, but the vast majority of them are not captives (Figure 3c). In this case, which is characteristic for the majority of cities in the developed world, the model (5) has three equilibria. A stable low equilibrium $B_{low}$, when the frequency of buses is low and only captive passengers use the line, a stable high equilibrium $B_{high}$, when the frequency of buses is high and the majority of potential passengers use the line, and an intermediate unstable equilibrium $B_{int}$, separating between the domains of attraction of towards the low and high equilibria. If the number of buses is below $B_{int}$, a typical vicious cycle starts; if it is above $B_{int}$, the system will enter a virtuous cycle as in the case presented in Figure 3d.

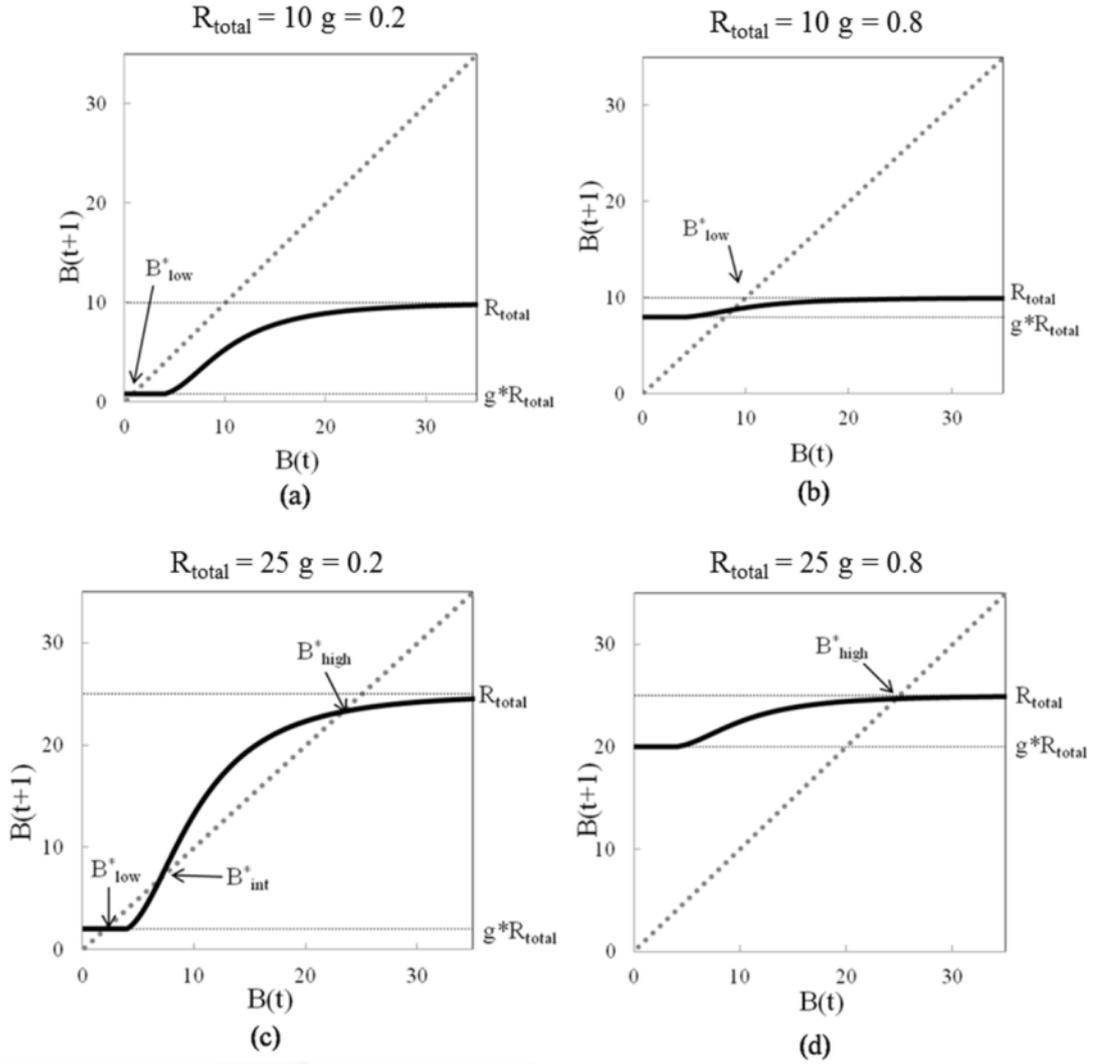

Figure 3: Four qualitatively different situations regarding the equilibria of (5), as dependent on $R_{total}$ and g, for m = 50 and for the distributions $f(\tau)$ of the willingness-to-wait according to Peterson et al. (2006). (a) $R_{total}$ = 10, g = 0.2; (b) $R_{total}$ = 10, g = 0.8; (c) $R_{total}$ =25, g=0.2; (d) $R_{total}$ = 25, g = 0.8.

The bifurcation diagram (Figure 4) represents the dependence of the equilibrium number of buses B*, according to (5), on the size of the population of potential passengers if the total population in the area served by the bus line is growing. The diagram is constructed assuming that the characteristics of the population – the share g of captive riders and the distribution of willingness-to-wait for a bus among non-captive riders – remain unaltered over the period of population growth.

If the majority of the population consists of captive riders (e.g., a "poor area" with a low level of car ownership), then the bus operator will react to the increase in the number of committed passengers by increasing bus frequency over time, thereby also inducing non-captive passengers to change their transportation mode in favor of the bus. This dynamics are reflected by the cross-section of the diagram for g = 0.8 (Figure 4c).

In case the majority of the population consists of non-captive passengers, the dynamics of the system are more complex. These dynamics are presented by the case of g = 0.2 (Figure 4b), when an initially low overall number of potential passengers (i.e., low $R_{total}$) enables one equilibrium state only, in which virtually only the captive passengers are served. With the increase in total population, the number of captive passengers will grow too and this will influence bus usage proportionally. However, the growth of the number of captive riders and the related increase in bus frequency is, in itself, insufficient for attracting a substantial number of non-captive riders and initiate virtuous cycle. The latter becomes possible after $R_{total}$ (for the values of g = 0.2 and m = 50) passes the lower threshold $R^1_{threshold} \approx 22$, when a second stable equilibrium B*$_{high}$ emerges. From this point onwards, the public transport operator can force the line into the virtuous cycle by instantaneously raising the number of buses above the unstable equilibrium B*$_{int}$ (the dotted line in Figure 4b). The situation changes once again when the total population size passes, in terms of number of buses, the higher threshold $R^2_{threshold} \approx 34$ (again, for the values of g = 0.2 and m = 50). From this point onwards, there is only one (high) equilibrium B*$_{high}$ (Figure 4b)

Note that within the interval $R^1_{threshold} < R_{total} < R^2_{threshold}$, the higher is $R_{total}$, the lower is the number of additional buses that should be added by the transport operator to initiate the virtuous cycle, and the higher the value of the second equilibrium B*$_{high}$ to which the system will converge after that (Figure 4b). At the same time, if the increase in the bus frequency is insufficient and the number of buses added by the operator would remain below B*$_{int}$, the operator will be forced back into the vicious cycle that will return the system to the low equilibrium B*$_{low}$.

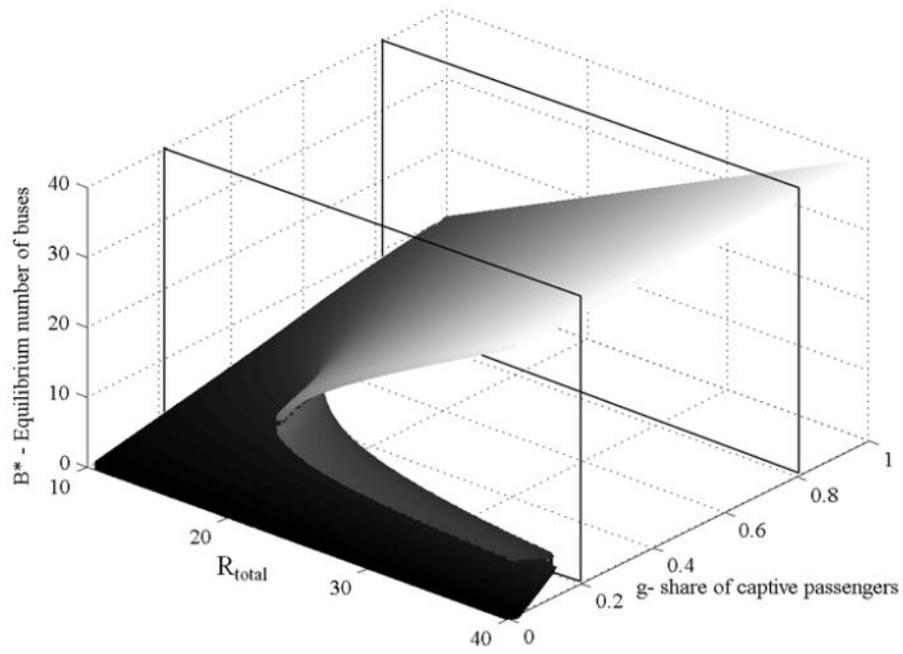

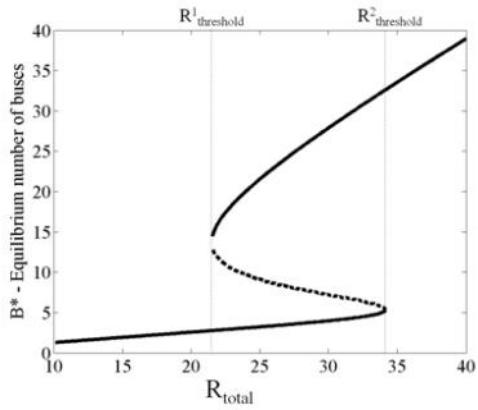
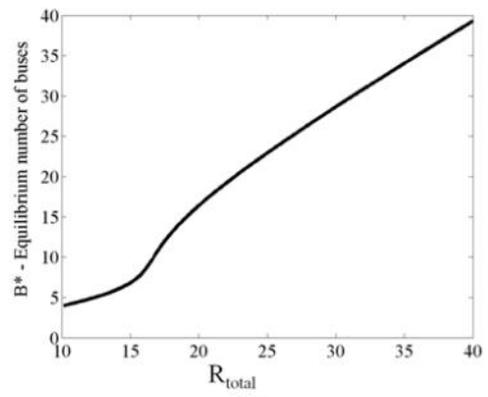

Figure 4: (a) Full bifurcation diagram of equation (4) for m = 50, varying $R_{total}$ and g, and the distribution of willingness-to-wait according to Peterson et al. (2006); (b) cross-section of the full diagram for low fraction of captive passengers, g = 0.2; (c) cross-section of the full diagram for high fraction of captive passengers, g = 0.8.

## 5. Extensions of the vicious cycle model

### 5.1. Bus size as a model parameter

One option for increasing bus frequency and, thus, initiating the virtuous cycle, is the use of smaller buses. Putting aside operation costs, the same total capacity of a bus line can be achieved by serving it with a higher number of smaller buses. In model terms, equation (4), where the size of a bus is reflected by the parameter m, is convenient for analyzing the impact of such an intervention. Just as above, if for a certain population size, the situation of one equilibrium (a line serving only captive passengers) is the only possible one (Figure 5a), then substitution of the line's large buses by more smaller buses, with smaller m, can turn the dynamics into a situation with two stable equilibria (Figure 5b) or even one (Figure 5c).

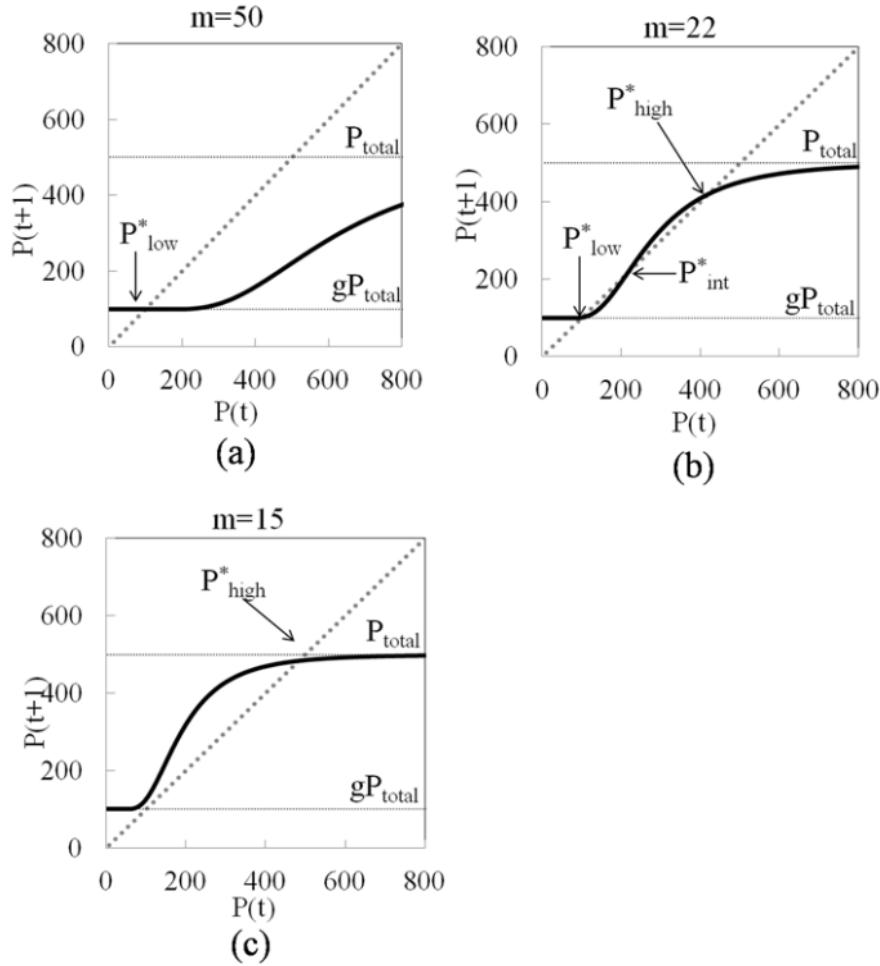

Figure 5: Evolution of model equilibria with a decrease in bus capacity m for $P_{total} = 500$ and $g = 0.2$: (a) m = 50, (b) m = 22, (c) m = 15.

The bifurcation diagram for the model (4) for $P_{total} = 1000$ and varying m and g, is presented in Figure 6.

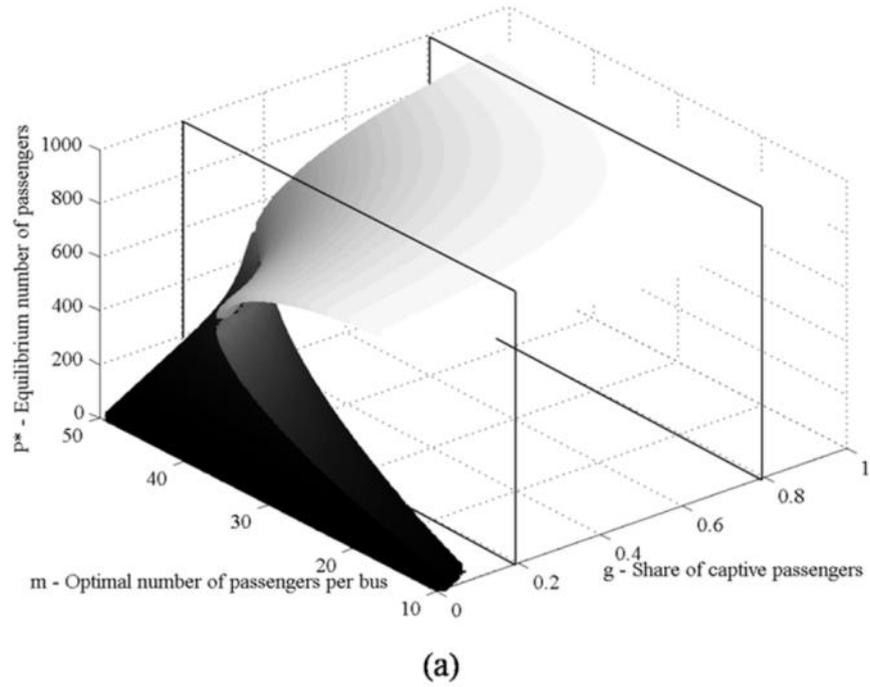

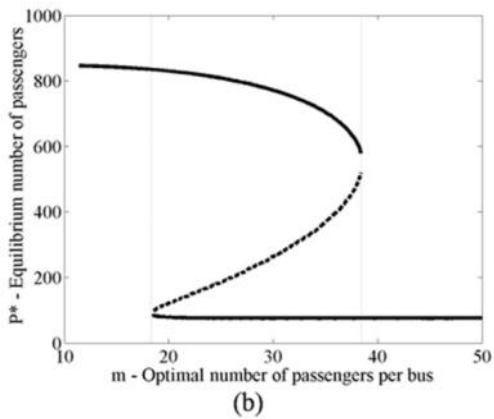 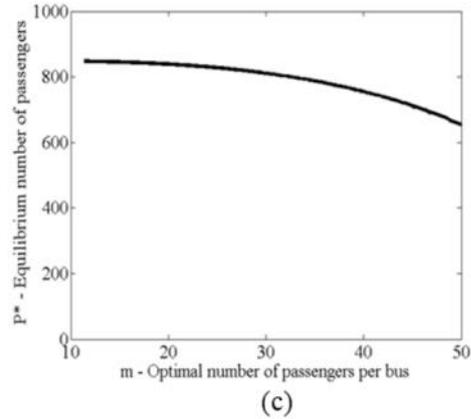

Figure 6: (a) Full bifurcation diagram of equation (5) for $P_{total} = 1000$, varying m and g, and distribution of willingness-to-wait according to Peterson et al. (2006); (b) cross-section of the full diagram for low fraction of captive passengers, $g = 0.2$; (c) cross-section of the full diagram for high fraction of captive passengers, $g = 0.8$.

### 5.2. One or two lines?

Let us consider establishing a bus service in a new area, assuming, for simplicity, that the spatial structure of the area enables a public transport service consisting of several circular lines only, in order to serve (parts of) the population in a reasonable way. Then, the basic question for the public transport operator is whether it is better – i.e., more profitable – to operate only one line that serves half of the population or to establish two lines that will serve twice as much potential passengers but with only half of the total number of buses per line?

Let us apply the model of the vicious cycle for the simplest case of an area that is characterized by uniform distributions of captive and non-captive passengers. Given a number of buses B and their size m, we compare one line of length L that serves $gP_{total}$ captives and $(1-g)P_{total}$ non-captive passengers by B buses to two non-related lines of length L, each serving $gP_{total}$ captive riders and $(1-g)P_{total}$ non-captive passengers with B/2 buses. Formally, the situation of two lines is identical to the situation of a line which length is 2L and which is served by the same number B of buses as the line of a length L.

In the first case, the interval between buses will be L/B, and the potential number of passengers $P_{total}$, while in the case of two lines, the interval between buses will be 2L/B, but with a potential number of passengers of $2P_{total}$. The dynamics of the number of passengers for each case, by quarters of a year, are given by:

$$P(t+1)_{1\ line} = gP_{total} + (1-g)P_{total} \int_{\frac{Lm}{P(t)}}^{\infty} f(\tau)d\tau \qquad (6)$$

$$P(t+1)_{2\ line} = 2gP_{total} + 2(1-g)P_{total} \int_{\frac{2Lm}{P(t)}}^{\infty} f(\tau)d\tau \qquad (7)$$

As can be seen directly from (6) – (7) in case of a majority of captive passengers, i.e. g ~ 1, two lines are preferable – in both cases $2gP_{total}$ passengers will be served, twice more than in case of one line, no matter what are the other parameters of the system.

However, in case of a non-captive majority, the situation is different: until the frequency of buses 2L/B is sufficient to pass the threshold that is necessary for entering the virtuous cycle, it is worth to put all buses into one line and not to serve the second line at all. For the intermediate values of g the situation becomes more complex: depending on the population density ($P_{total}$/L), the number of equilibria for one line of length L can be different form the number of equlibria of the two-line situation. We delay the study of this case to future papers.

## 6. Conclusion and discussion

In this paper, we have presented an analytical model for analyzing the vicious and virtuous cycles in public transport based on passengers' decisions to use a public transport service dependent on waiting time. While based on a number of simplifying assumptions – a single line instead of a transportation network, a uniform distribution of captive and non-captive passengers in space and over the day and zero operation costs - the model shows that service provision will either enter a vicious or virtuous cycle. The dynamics of the vicious cycle converges to an equilibrium that is characterized by low service frequency and low ridership, while the dynamics of the virtuous cycle converges to equilibrium of high service frequency and high ridership. We have analyzed the conditions of the vicious or virtuous cycle based on available field data regarding passengers' variation in willingness-to-wait for a public transport service. The analysis substantially adds to the discussion by Levinson and Krizek (2008) who used hypothetical dependencies between the frequency of buses and the number of passengers. In future work, we aim to address the issues of non-uniform distribution of passengers over space, variation in trip lengths and real-time travel information provided through smart phones.

Despite intentionally simplifying assumptions, we argue that a number of practical lessons can be drawn from the analysis of our abstract model. Let us address two situations that typify a large part of the world: city regions in wealthy countries and city regions in rapidly developing but relatively poor countries.

Typically, city regions in wealthy countries like the US, Japan, or in Western-European countries, are characterized by low shares of captive riders. Total number of potential passengers in these city regions depends highly on urban structure. In low density sprawling areas, public transport lines, if still existent, will inevitably enter a vicious cycle if subsidies are terminated (Figure 3a). In denser areas, with a clustering of employment and other activities in a limited number of centers, the possible line dynamics are probably best described by Figure 3c and both a vicious and a virtuous cycle may be possible:

- Frequencies on many public transport lines will be below the $B^*_{int}$ level and so are potentially in a vicious cycle. The level of service on these lines will currently be maintained by subsidies. However, as the share of captive riders drops over time due to increasing real incomes and related car ownership, these lines will keep facing a decreasing ridership and thus subsidies will inevitably have to increase. This may not be possible due to budget limitations, forcing a further decrease in service. In real-world circumstances with limited budgets, a vicious cycle therefore seems inevitable, even if the public transport service is fully regulated. Eventually, these lines will converge to a politically acceptable minimal level of service or will be terminated altogether.
- Some lines may actually be close to the $B^*_{int}$ level. For these lines, a temporary increase in subsidy combined with a requirement to increase the bus frequency may lead to a virtuous cycle. If subsidies are maintained long enough, the line, through the virtuous cycle, would converge to high equilibrium. Note that subsidies could be reduced (substantially) at the moment the high equilibrium is reached, depending on operating costs and fare box revenues, suggesting that short-term increases in subsidies may actually lead to a better level-of-service and lower annual subsidies in the long-run.
- Few lines may already be above the $B^*_{int}$ level, but be part of integrated subsidized tenders. In such cases, the public transport operator has no incentive to increase the level of service on these lines. It may be considered to take these lines out of the tenders and to allow operation on a commercial basis only, to trigger public transport operators to make better use of the potential offered by the virtuous cycle.

Other lines may benefit from this split, both from the increase in per line subsidy and, through a network effect, from a total increase in ridership.

Recent social shifts, such as a decreasing car ownership among young adults and a leveling out of car mobility in a number of Western countries, suggest that high-density cities may actually experience an increase in the share of (voluntarily) captive passengers over the coming decade. This can be a trigger for a virtuous cycle on an increasing number of public transport lines. The experiences in Berlin, as reported by Reinhold (2008), show that it is indeed possible to 'engineer' a virtuous cycle for particular bus lines by increasing service frequency. However, the Berlin bus system is still heavily subsidized and service frequency is determined top-down rather than in direct response to passenger demand. It thus remains unclear whether the virtuous cycle for the main bus lines there is due to sustained subsidies or whether these lines could maintain a high equilibrium without government support.

Overall, our analysis suggests that in wealthy countries it might be attractive, both in terms of ridership and total subsidy needs, to change from area-based tenders of public transport service (Farsi et al., 2007) to a system that makes a distinction between bus lines vis-à-vis the vicious/virtuous cycle, in spite of the possible drawbacks of such a system.

Urban public transport services in the emerging economies are often in a completely different situation. They are usually characterized by a large potential passenger population and a large share of captive riders. However, several developments suggest that these cities may rapidly move into a situation in which a vicious cycle could occur. First, these cities are experiencing a rapid increase in car ownership. Second, operating costs and therefore ticket prices are likely to increase over the coming decade, due to higher demand for quality of vehicles/services and increasing labor costs. Taken together, these development suggest a decreasing share of captive riders and a lower use of public transport among non-captive riders, which may in turn induce operators to reduce frequencies, with a risk of entering the vicious circle. The challenge for these cities will be to identify impending vicious cycles and prevent them from happening through 'early-bird' subsidies and/or strategies to reduce operating costs and increase service frequencies, such as free bus lanes and traffic light priority, in order to return to a path of a virtuous cycle. If successful, this may also reduce the rate with which the captive population will turn into a non-captive one.

A final observation regards the importance of the use of models for studying the possible development of the transportation system over time. Currently, virtually all transportation-planning agencies, whether dealing with public and/or private transport, use static and aggregate models to analyze the dynamics of real-world transport systems. The models are static in nature, in that they assume that future travel demand can be forecasted largely independently of future transport infrastructure or transport policies. The models are aggregate in that they do not simulate the behavior of the individual traveler. As such, these models are insufficient for simulating the interrelationships between the users of the transport system, the impacts of policies over time, and the intricate relation between short-and long-term dynamics of the transport system. Static aggregate transport models tend to generate results that satisfy predicted demand, but cannot identify possibilities to change the system through measures which impact becomes only apparent over time. Our study underscores that high-resolution dynamic models may substantially enrich transportation modeling and, in its wake, transportation policies.

# Appendix A

Let us analytically investigate equilibria of the equation

$$B(t+1) = F(B(t)) = R_{total}[g + (1-g)\int_{\frac{L}{B(t)}}^{\infty} f(\tau)d\tau] \quad (A1)$$

The first derivative of the F(B(t)) by B is given by

$$\frac{dF}{dB} = R_{total}(1-g)L \times \frac{f\left[\frac{L}{B}\right]}{B^2} \quad (A2)$$

and is positive for all B, because f(τ) is always positive. That is, F(B) monotonously grows with the growth of B. In case f(τ) is monotonous, equilibria of (A1) and their stability can be fully investigated analytically. Indeed, for monotonous f(τ), equilibria and their stability is defined by whether the function F(B(t)) is concave or convex. The latter is defined by the second derivative of the F(B):

$$\frac{d^2F}{dB^2} = R_{total}(1-g)L\left[-\frac{2f\left[\frac{L}{B}\right]}{B^3} - \frac{lf'\left[\frac{L}{B}\right]}{B^4}\right] \quad (A3)$$

F(B) is concave if the second derivative, given by (A3), is positive and convex if it is negative. The sign of the second derivative is defined by $-2f\left[\frac{L}{B}\right] - \frac{L}{B}f'\left[\frac{L}{B}\right]$ or, substituting L/B by $T_B$, by

$$-f'[T_B] - \frac{2f[T_B]}{T_B} \quad (A4)$$

We can now specify monotonous function f(τ) that separates between concave and convex F(B(t)). Solving differential equation $0 = -f'[\tau] - \frac{2f[\tau]}{\tau}$ we obtain

$$f[\tau] = \frac{C}{\tau^2} \quad (A5)$$

where C is a constant. Assuming that τ varies on [$\tau_{min}, \tau_{max}$] and applying $\int_{\tau_{min}}^{\tau_{max}} f(\tau)d\tau = 1$ we obtain $C = \frac{\tau_{max} \times \tau_{min}}{\tau_{max} - \tau_{min}}$.

Based on (A5), three options of the dynamics of (A1) for monotonous f(τ) are interesting (Figure A1):

1. For f(τ) = C/τ², F(B) is a straight line. In case of high $R_{total}$, the bus line converges to a state in which all potential passengers use it. In case of low $R_{total}$, the system converges to a stable equilibrium ($F'(B^*) < 1$) (Figure A1b).
2. For f(τ) that decreases slower than C/τ², e.g., f(τ) = C/τ, the system always converges to a single equilibrium ($F'(B^*) < 1$), in which almost all passengers use the bus line (Figure A1a).
3. For f(τ) that decreases faster than C/τ², e.g. f(τ) = C/τ³, one equilibrium is characteristic of the system in case of low $R_{total}$, and two in case of high $R_{total}$. The single or the lower of the two equilibria is always stable. The higher of two equilibria is always unstable (Figure A1c).

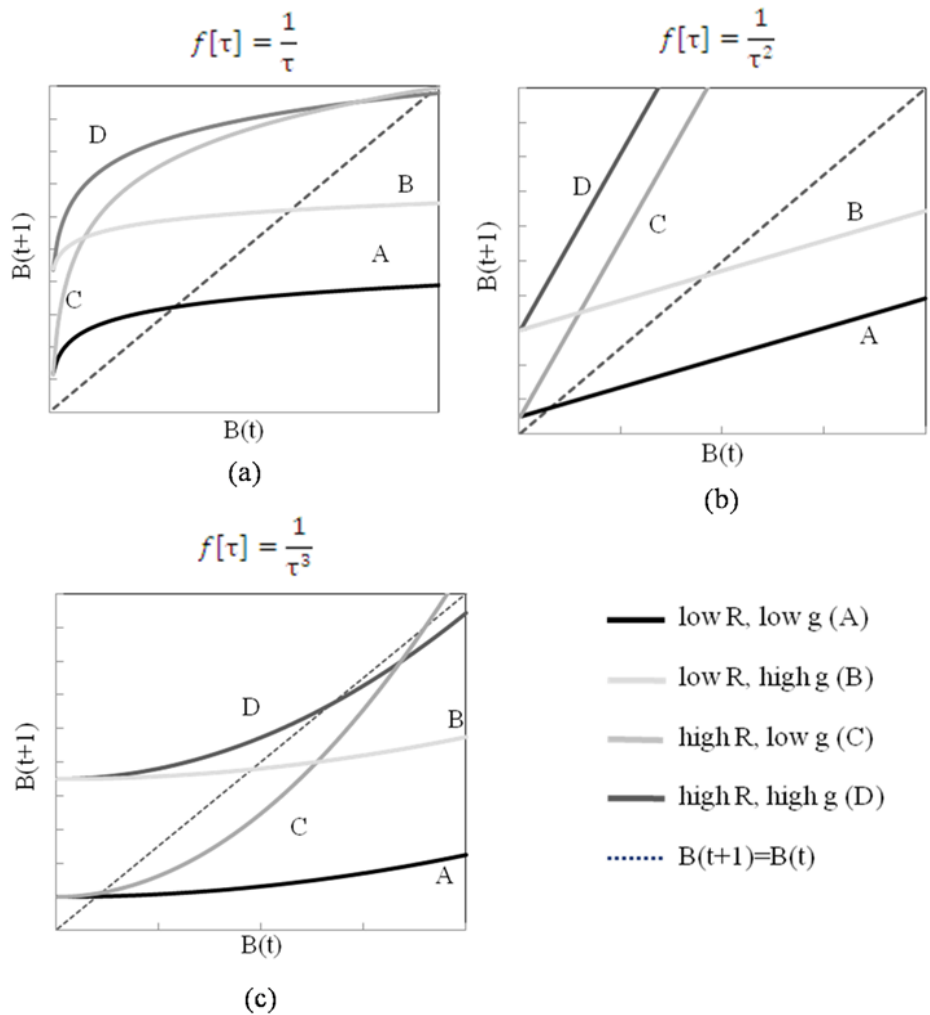

Figure A1. The shape of F(B) in case of $f(\tau) = C/\tau$, $f(\tau) = C/\tau^2$ and $f(\tau) = C/\tau^3$. The four curves A-D in each figure differ in terms of the fraction of captive passengers g (low/high) and overall number of potential passengers $R_{total}$ (low/high): (A) $R_{total} = 20$, $g = 0.2$, (B) $R_{total} = 20$, $g = 0.8$, (C) $R_{total} = 70$, $g = 0.2$, and (D) $R_{total} = 70$, $g = 0.8$.

To conclude, monotonous $f(\tau)$ always results in the vicious cycle in case $f(\tau)$ decreases, with $\tau$, faster than $C/\tau^2$, and always results in the virtuous cycle in case $f(\tau)$ decreases, with $\tau$, slower than $C/\tau^2$.

# Appendix B

If $\tau < T_B$ then

$$N_u(\tau, d+1) = N_u(\tau, d) + \alpha N_n(\tau, d)\frac{\tau}{T_B} - \beta N_u(\tau, d)\left(1 - \frac{\tau}{T_B}\right) \qquad (B1)$$

$$N_n(\tau, d+1) = N_n(\tau, d) - \alpha N_n(\tau, d)\frac{\tau}{T_B} + \beta N_u(\tau, d)\left(1 - \frac{\tau}{T_B}\right)$$

Let:

$$\alpha\frac{\tau}{T_B} = \epsilon;\ \beta\left(1 - \frac{\tau}{T_B}\right) = \delta;\ N_u(\tau, d) = X_t;\ \text{and}\ N_n(\tau, d) = Y_t$$

Equation (B1) can be thus presented as:

$$X_{t+1} = X_t(1 - \delta) + \epsilon Y_t \qquad (B2)$$
$$Y_{t+1} = \delta X_t + Y_t(1 - \epsilon)$$

And in a matrix form as:

$$\begin{pmatrix} X_{t+1} \\ Y_{t+1} \end{pmatrix} = \begin{pmatrix} 1 - \delta & \epsilon \\ \delta & 1 - \epsilon \end{pmatrix} \begin{pmatrix} X_t \\ Y_t \end{pmatrix}$$

To find the eigenvalues of matrix $\begin{pmatrix} 1 - \delta & \epsilon \\ \delta & 1 - \epsilon \end{pmatrix}$ we have to solve the equation:

$$\lambda^2 - \lambda(2 - \delta - \epsilon) + (1 - \delta - \epsilon) = 0$$

Let us denote $(1 - \delta - \epsilon)\ as\ a$; the roots of (3) are thus

$$\lambda_{1,2} = \frac{1 + a \pm \sqrt{(1+a)^2 - 4a}}{2} = \frac{1 + a \pm (1 - a)}{2}$$

That is,

$$\lambda_1 = 1$$

and

$$\lambda_2 = a = 1 - \delta - \epsilon = 1 - \beta + \frac{\tau}{T_B}(\beta - \alpha) < 1$$

The solution of equation (B2) thus converges, in days, to the eigenvector corresponding to the largest eigenvalue $\lambda_1 = 1$. The components of this eigenvector are as follows:

$$X^* = \frac{N_0 \epsilon}{\epsilon + \delta},\quad Y^* = \frac{N_0 \delta}{\epsilon + \delta}$$

That is, for given $\tau < T_B$, the ratio between the number of potential passengers who use the bus and the number of potential passengers who do not use the bus, converges, in days, to

$$\frac{N^*u(\tau)}{N^*n(\tau)} = \frac{X^*(\tau)}{Y^*(\tau)} = \frac{\epsilon}{\delta} = \frac{\alpha}{\beta}\frac{\tau}{(T_B - \tau)}$$

For τ < $T_B$, the equilibrium fraction of bus users characterized by willingness-to-wait τ is, thus, equal to:

$$u(\tau) = f(\tau)\frac{1}{1+\frac{\beta}{\alpha}\frac{(T_B-\tau)}{\tau}} \qquad (B3)$$

Figure 2, in the main text, shows the effect of α and β on the number of users at the equilibrium.

The number of bus users at d = 90 that are characterized by the willingness–to-wait τ < $T_B$ is thus given by:

$$P(t+1)_{\tau<T_B} = N\int_0^{T_B} f(\tau)\frac{1}{1+\frac{\beta}{\alpha}\frac{(T_B-\tau)}{\tau}}d\tau$$

and depends on the analytical expression of f(τ).

For example, in case f(τ) is uniform on [0, $T_{max}$]

$$P(t+1)_{\tau<T_B} = N\int_0^{T_B} \frac{C}{1+\frac{\beta}{\alpha}\frac{(T_B-\tau)}{\tau}}d\tau$$

Where C = 1/$T_{max}$ and, analytically,

$$P(t+1)_{\tau<T_B} = C*N*T_B*F\left(\frac{\beta}{\alpha}\right), \text{ where } f(x) = \frac{x[log(x)-1]+1}{(x-1)^2}.$$

According to (B3), the full model of a bus line is, thus:

$$P(t+1) = A + N\left[\int_0^{T_B(t)} \frac{f(\tau)}{1+\frac{\beta(T_B(t)-\tau)}{\alpha}\frac{}{\tau}}d\tau + \int_{T_B(t)}^{\infty} f(\tau)\,d\tau\right] \qquad (B4)$$

and it differs from the model in the main text by the additional (first) term in the square brackets. This addition, however, does not change in any qualitative way the dynamics of the system. Figure B1 repeats Figure 3c in the main text for high $R_{total}$, low g, α = 0.005, and β = 0.50, 0.25 and 0.05 per day (i.e., $\frac{\beta}{\alpha}$ = 100, 50, 10). The only significant effect, as can be expected, is the increase in the low equilibrium $B_{low}$ with the decrease in $\frac{\beta}{\alpha}$.

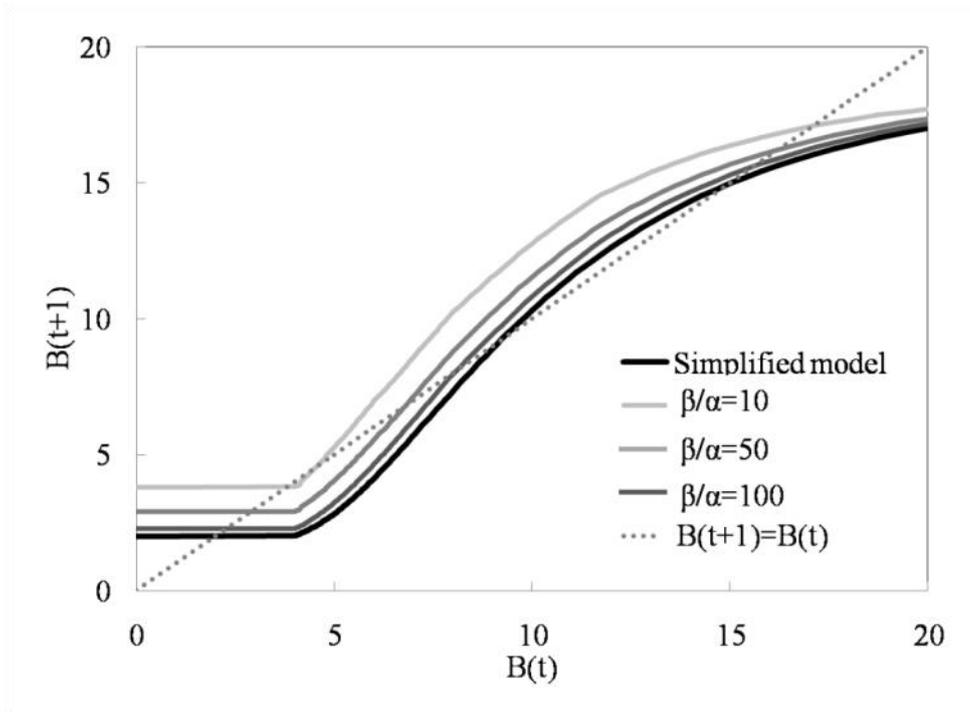

Figure B1: Dependence of B(T + 1) on B(T) as expressed by (B4), for the distribution of willingness-to-wait according to Peterson et al. (2006), $R_{total} = 20$, $g = 0.1$, and three different values of $\frac{\beta}{\alpha}$. The black curve represents the same dependency for the simplified model that is investigated in the main text (Figure 3c).

# Appendix C: Analysis of Levinson and Krizek's (2008) model

The model presented by Levinson and Krizek (2008) is based on three heuristic equations that relate between bus speed v, number of passengers P, and bus waiting time W. They do not specify in their analysis which of the equations is chosen for performing the time-step transition and we have arbitrarily chosen their second equation, as presented in (C2) below:

$$v_t = 1.5 - 0.5 W_t \quad (C1)$$

$$\frac{P_{t+1}}{3000} = \frac{e^{v_t}}{1+e^{v_t}} \quad (C2)$$

$$W_t = 30 \left(\frac{1}{0.5+0.02*P_t}\right) \quad (C3)$$

To analyze the system we express $P_{t+1}$ as a function of $P_t$, substituting (C1) into (C2):

$$P_{t+1} = 3000 \frac{e^{1.5-0.5W_t}}{1+e^{1.5-0.5W_t}}$$

and, further, substituting (C3) into the result:

$$P_{t+1} = F(P_t) = 3000 * \frac{e^{1.5-\frac{750}{25+P_t}}}{1+e^{1.5-\frac{750}{25+P_t}}} \quad (C4)$$

Dependency (C4) is presented in Figure C1b, while Figure C1a presents a zoom of the coordinate plane for the values of $P_t$ and $P_{t+1}$ between 0 and 500. As can be seen, the equations of Levinson and Krizek (2008) produce three equilibria as is characteristic for a high R and low g, but result in non-realistic values of $B_{int} \approx 150$ and $B_{high} \approx 1,200,000$.

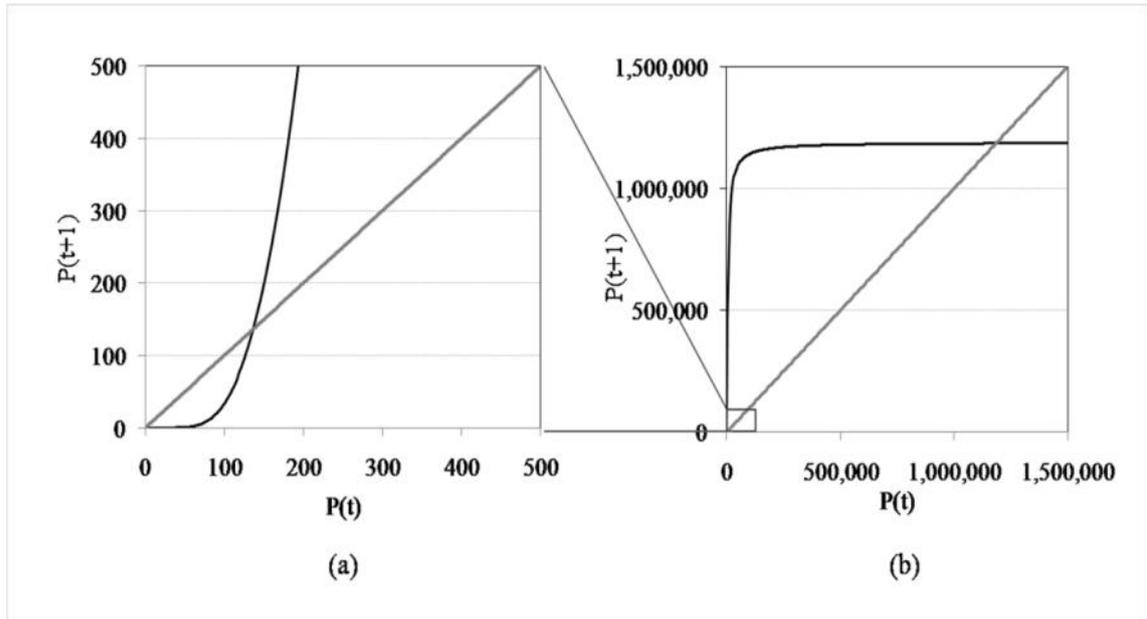

Figure C1: Equilibria of Levinson and Krizek's (2008) model: (a) dependence of P(T + 1) on P(T) for low values of P(T); and (b) dependence of P(T + 1) on P(T).

# 7. References

Bly, P., 1987. Managing public transport: Commercial profitability and social service. *Transportation Research Part A: General* 21, 109-125.

Castaline, A., 1980. Service Evaluation Guidelines. Office of Planning Assistance, UMTA, US Department of Transportation, Washington DC.

Caulfield, B., O'Mahony, M., 2009. A stated preference analysis of real-time public transit stop information. *Journal of Public Transportation* 12, 1-20.

Ceder, A., 2007. *Public transit planning and operation: theory, modelling and practice*. Butterworth-Heinemann, Oxford/Burlington.

Downs, A., 1962. The law of peak-hour expressway congestion. *Traffic Quarterly* 16.

Farsi, M., Fetz, A., Filippini, M., 2007. Economies of Scale and Scope in Local Public Transportation. *Journal of Transport Economics and Policy* 41, 345-361.

Furth, P., Muller, T., 2006. Part 4: Capacity and Quality of Service: Service Reliability and Hidden Waiting Time: Insights from Automatic Vehicle Location Data. *Transportation Research Record: Journal of the Transportation Research Board* 1955, 79-87.

Hess, D.B., Brown, J., Shoup, D., 2004. Waiting for the Bus. *Journal of Public Transportation* 7, 67-84.

Holmgren, R.A., 1996. *A first course in discrete dynamical systems*, Second edition ed. Springer Verlag, New York/Berlin/Heidelberg.

Kim, Y., Ceder, A., 2006. Smart feeder/shuttle bus service: consumer research and design. *Journal of Public Transportation* 9.

Kingham, S., Dickinson, J., Copsey, S., 2001. Travelling to work: will people move out of their cars. *Transport Policy* 8, 151-160.

Levinson, D.M., Krizek, K.J., 2008. *Planning for place and plexus: metropolitan land use and transport*. Taylor & Francis, NewYork/London.

Liu, S., Triantis, K.P., Sarangi, S., 2010. A framework for evaluating the dynamic impacts of a congestion pricing policy for a transportation socioeconomic system. *Transportation Research Part A: Policy and Practice* 44, 596-608.

Martens, K., Hurvitz, E., 2011. Distributive impacts of demand-based modelling. *Transportmetrica* 7, 181-200.

Peterson, D., Hough, J., Ulmer, D., 2006. Express Bus Transit Study: A Case Study. Upper Great Plains Transportation Institute/North Dakota State University, Fargo.

Reinhold, T., 2008. More Passengers and Reduced Costs—The Optimization of the Berlin Public Transport Network. *Journal of Public Transportation* 11.

Van Nes, R., 2002. *Design of multimodal transport networks: A hierarchical approach* IOS Press, Delft.

Xu, M., Ceder, A., Gao, Z., Guan, W., 2010. Mass transit systems of Beijing: governance evolution and analysis. *Transportation* 37, 709-729.